# Synthetic Multiferroic Interconnects for Magnetic Logic Circuits

Alexander Khitun

Electrical Engineering Department, University of California Riverside, California 92521

**Abstract**

In this work, we consider the possibility of using synthetic multiferroics comprising piezoelectric and magnetostrictive materials as an interconnect for nano-magnetic logic circuits. The proposed interconnect resembles a parallel plate capacitor filled with a piezoelectric, where one of the plates is made of a magnetoelastic material. The operation of the interconnect is based on the effect of stress-mediated anisotropy modulation, where an electric field applied across the piezoelectric material produces stress, which, in turn, affects the anisotropy field in the magnetostrictive material. We present the results of numerical modeling illustrating signal propagation through the interconnect. The model combines electric and magnetic parts, where the electric part describes the distribution of an electric field through the piezoelectric and the magnetic part describes the change of magnetization in the magnetoelastic layer. The model is based on the Landau–Lifshitz–Gilbert equation with the electric field dependent anisotropy term included. The utilization of the electro-magnetic coupling makes it possible to amplify magnetic signal during its propagation via energy conversion from the electric to magnetic domains. Potentially, synthetic multiferroic interconnects can be implemented in a variety of spin-based devices ensuring reliable and low-energy consuming data transmission. According to the estimates, the group velocity of magnetic signals may be up to $10^5$ m/s with energy dissipation less than $10^{-18}$ J per bit per 100nm. The fundamental limits and practical shortcoming of the proposed approach are also discussed.

I.      Introduction

There is a growing interest in novel computational devices able to overcome the limits of the current complimentary-metal-oxide-semiconductor (CMOS) technology [1]. Magnetic logic circuits are among the most promising approaches offering a significant reduction of consumed power by utilizing the inherent non-volatility of magnetic elements. In magnetic logic circuitry, a bit of information is encoded into the magnetization state of a nano-magnet, which may be kept for long time without any power consumption, while the external energy is required only to perform computation (i.e. switching between the magnetization states). Though magnetic memory became a widely used commercial product a long time ago, magnetic logic is pretty much at its infancy stage. The development of energetically efficient and reliable magnetic interconnects is one the main challenges to be solved. Similar to electronic transistor-based circuits, where one transistor drives the next stage transistors by electric signals, magnetic logic circuits requires one magnet to drive the next stage magnets by sending magnetic signals. There are different possible ways to interconnect the input and the output magnets (i.e. by making an array of nano-magnets sequentially switched in a domino fashion[2], by sending a spin polarized current[3], by sending a spin wave[4], or by moving a domain wall [5]). There is always a tradeoff between the speed, the energy per bit and the reliability of magnetic signal transmission. It takes either a large amount of energy for error-prone signal transmission or the error probability increases with the distance due to the thermal noise, defects, and signal dispersion. The lack of amplification is one of the key issues inherent to the above mentioned approaches. In this work, we consider synthetic multiferroics for magnetic interconnects, which may provide magnetic signal amplification by transferring energy between the electric and magnetic domains.

Synthetic multiferroics (or two-phase composite multiferroics) comprise piezoelectric and magnetoelastic materials, where an electric field applied across the piezoelectric produces stress, which, in turn, affects the magnetization of the magnetoelastic material. Although work in this area can be traced back to the 1970s [6], synthetic multiferroics have been in a shadow of the single-phase multiferroics (i.e. $BiFeO_3$ and its derivatives[7]) for a long time. The recent resurgence of interest in composite multiferroics is due to the technological flexibility allowing for independent variation of piezoelectric or magnetostrictive layers. More importantly, the strength of the electro-magnetic coupling in the two-phase systems can significantly exceed the limits of the single-phase counterparts [8]. During the past two years, there have been several experimental works showing magnetization rotation in two-phase multiferroics as a function of the applied voltage[9,10]. For instance, a reversible and permanent magnetic anisotropy reorientation was reported in a magnetoelectric polycrystalline Ni thin film and (011)-oriented $[Pb(Mg_{1/3}Nb_{2/3})O_3]_{(1-x)}-[PbTiO_3]_x$ (PMN-PT) heterostructure [9]. An important feature of the magneto-electric coupling is that the changes in magnetization states are stable without the application of an electric field and can be reversibly switched by an electric field near a critical value (i.e. 0.6 MV/m for Ni/PMN-PT). Such a relatively weak electric field promises an ultra-low energy consumption required for magnetization rotation[11]. The idea of using stress-mediated mechanism for nano-magnet switching is currently extensively studied [12] [13]. Here, we propose to extend this approach to magnetic interconnects and exploit the strain mediated electro-magnetic coupling for magnetic signal amplification. The rest of the paper is organized as follows. In Section II, we describe the material structure and the principle of operation of the synthetic multiferroic interconnects. The results of numerical modeling illustrating signal propagation are presented in Section III. The Discussion and Conclusions are given in Sections IV and V, respectively.

## II. Material Structure and Principle of Operation

The schematics of the proposed interconnect on top of a silicon wafer are shown in Figure 1(A). It consists from the bottom to the top from a conducting layer (e.g. Pt), a layer of piezoelectric material (e.g. PMN-PT), and a layer of magnetoelastic material (e.g. Ni). The whole structure represents a parallel plate capacitor filled with a piezoelectric, were one plane (the bottom) is made of a non-magnetic metal and the top plate is made of a magnetoelastic metal. The top layer is the media for magnetic signal propagation between the nano-magnets to be placed on the top of the layer. For simplicity, we have shown just two nano-magnets marked as A and B in Figure 1(A). The nano-magnet market A is the input element to send a magnetic signal to the receiver nano-magnet B. The spins of the nano-magnets are coupled to the spins of the ferromagnetic magnetostricitve layer via the exchange interaction. The nano-magnets are assumed to be of a special shape to ensure the two thermally stable states of magnetization. Hereafter, we assume the magnetoelastic layer to be polarized along the X axis, and the nano-magnets to have two states of magnetization along or opposite to the Y axis. Each of the nano-magnets has an electric contact where a control voltage is applied. The bottom layer made of a nonmagnetic metal serves as a common ground plate.

The principle of operation is the following. In order to send a signal from A to B, a control voltage V is applied to the nano-magnet A. The applying of voltage starts the charge diffusion through the conducting plates. The equivalent circuit is shown in Figure 1(B). The charge diffusion through the capacitor plates is well described by the RC model, where the resistance R and the capacitance C are defined by the geometric size and the material properties of the conducting plates and the piezoelectric layer. An electric field appearing across the piezoelectric produces stress, which affects the anisotropy of the magnetostrictive material by rotating its easy-axis. It is assumed that the applying of voltage rotates the easy axis from X axis towards the Y axis. The change of the anisotropy field caused by the applied voltage affects the magnetization of the magnetoelastic layer. There are two possible trajectories for the magnetization to follow: along or opposite to the Y axis. The particular trajectory is defined by the magnetization state of the sender nano-magnet A. (i.e. the magnetization of the ferromagnetic layer copies the magnetization of the sender nano-magnet).

In Figure 1(C), we present the results of numerical modeling showing the snapshot of the distribution of the electric field E(x), and the magnetization component My(x) through the interconnect. The details of numerical modeling are presented in the next Section. Here, we want to illustrate the main idea of using synthetic multiferroics as a magnetic interconnect: magnetic signals (i.e. the local change of magnetization) can be sent through large distances without degradation as the angle of magnetization rotation is controlled by the applied voltage. The direction of signal propagation (e.g. from A to B, or vice versa) is controlled by the applied voltages as well. The charging of the capacitor eventually leads to the uniform electric field distribution among the plates and static distribution of magnetization through the magnetoelastic layer. There are several possible ways for the output nanomagnet B switching. For example, it can be preset in a metastable state prior to computation (e.g. magnetization along the Z axis), so the magnetic signal sent by A triggers the relaxation towards the one of the thermally stable states along or opposite to the Y axis. There may be also possible scenarios where the receiver nano-magnet is connected to the two or more nano-magnets, so the final state is defined by the interplay of several incoming signals (e.g. MAJ operation). In this work, we focus on the mechanism of signal transmission only, though the utilization of synthetic multiferroic interconnects may further evolve the design of magnetic logic circuits similar to ones presented in Refs. [3,4,14].

### III. Numerical Modeling

The model for signal propagation in the synthetic multiferroic combines electric and magnetic parts. The electric part is aimed to find the distribution of an electric field through the piezoelectric, and the magnetic part describes the change of magnetization in the magnetoelastic layer. The charge distribution is modeled via the following equation:

$$R_s C_s \frac{d^2 V(x,t)}{dx^2} = \frac{dV(x,t)}{dt} \quad , \tag{1}$$

where $R_s$ and $C_s$ are the resistance and capasitance per unit length, V(x,t) is the voltage distribution over the distance. The simulations start with V(0,0)=$V_{in}$, and V(x,0)=0 everywhere else thorugh the plates.

The process of the magnetization rotation is modeled via the Landau-Lifshitz equation:

$$\frac{d\vec{m}}{dt} = -\frac{\gamma}{1+\eta^2} \vec{m} \times \left[ \vec{H}_{eff} + \eta \vec{m} \times \vec{H}_{eff} \right] \quad , \tag{2}$$

where $\vec{m}$ is the unit magnetization $\vec{m} = \vec{M}/M_s$ vector, $M_s$ is the saturation magnetization, $\gamma$ is the gyro-magnetic ratio, and $\eta$ is the phenomenological Gilbert damping coefficient. The effective magnetic field $\vec{H}_{eff}$ is the sum of the following:

$$\vec{H}_{eff} = \vec{H}_d + \vec{H}_{ex} + \vec{H}_a + \vec{H}_b \tag{3}$$

where $H_d$ is the magnetostatic field, $H_{ex}$ is the exchange field , $H_a$ is the anisotropy field $\vec{H}_a = (2K/M_S)(\vec{m} \cdot \vec{c})\vec{c}$ (K is the uniaxial anisotropy constant, and $\vec{c}$ is the unit vector along the uniaxial direction), $H_b$ is the external bias magnetic field. The two parts are connected via the voltage-dependent anisotropy term as follows:

$c_x$=cos($\theta$), $c_y$=sin($\theta$), $c_z$=0 (4)

$$\theta = \frac{\pi}{2}\left(\frac{V(x)}{V_\pi}\right)$$

where $V_\pi$ is the voltage resulting in a 90 degree easy axis rotation in the X-Y plane.

The introduction of the voltage-dependent anisotropy field (Eq.4) significantly simplifies simulations as it presumes an immediate anisotropy response on the applied electric field without considering the stress-mediated mechanism of the electro-magnetic coupling. Such a model can be taken as a first-order approximation. Nevertheless, this model is useful in capturing the general trends of signal propagation and can provide estimates on the maximum speed of signal propagation and energy losses. In our numerical simulations, we use the following material parameters: the dielectric constant ε of the piezoelectric is 2,000; the electrical resistivity of the magnetoelastic material is $7.0 \times 10^{-8}$ Ω·m, the gyro-magnetic ratio $\gamma = 2 \times 10^7$ rad/s, the saturation magnetization Ms=10kG/4π; $2K/M_s$=100Oe, external magnetic field $H_b$=100Oe is along the X axis, and the Gilbert damping coefficient $\eta$=0.1 for the magnetostrictive material. For simplicity, we also assumed the same resistance for the bottom and the top conducting plates. The strength of the electro-magnetic coupling (i.e. $V_\pi$) is calculated based on the available experimental data for PMN-PT/Ni (i.e. 0.6 MV/m for 90 degree rotation [9]).

The results of numerical simulations shown in Figure 1 (C) are obtained for the interconnect comprising 40nm of piezoelectric and 4nm of magnetoelastic materials. The two curves in Figure 1(C) depict the distribution of the electric field E(x) and the projection of magnetization My(x) along the interconnect after the voltage has been applied through the nano-magnet A. The curves are plotted in the normalized units $E/E_0$ and $M_y/M_s$, where $E_0=V_\pi/d$, were d is the thickness of the multiferroic layer (40nm). The distribution of the electric filed was found by solving Eq.(1). Then, the anisotropy field was found via Eq. (4), and, finally, magnetization change was simulated via Eqs. (2-3). The results in Fig.1(C) show a snapshot taken at 0.4ns after the voltage has been applied. In these simulations, we assumed the nano-magnet A to be polarized along the Y axis, and the magnetization of the interconnect beyond the nano-magnet $M_y(0)$=0.1$M_s$ due to the exchange coupling with the spins of the nano-magnet. The spins of the magnetoelastic material are tend to rotate in the same direction as the spins of the sender nano-magnet A. Eventually, the Y-component of magnetization of the interconnect saturates along the constant value, which is defined by the interplay of the anisotropy and the bias magnetic fields.

In Figure 2, we show the results of numerical modeling illustrating the dynamics of magnetization rotation in the interconnect. The curves in Figure 2 depict the evolution of local magnetization in the interconnect located 1μm, 2μm and 3μm away from the excitation point. The insets in Fig.2 show the initial state of magnetization of the sender nano-magnet A. In all cases, the magnetization trajectory in the interconnect repeats the initial magnetization of the nano-magnet A (e.g. the magnetization component My is positive if nano-magnet A is polarized along the Y axis, and the $M_y$ is negative if nano-magnet A is polarized opposite to the Y axis). The absolute value of the final steady state is the same (about 0.5$M_s$) for all six curves. These results illistrate the main idea of implementing electric field-driven multiferroic interconnect allowing to maintain the amplitude of the magnetic signal constant regardless the propagation distance.

## IV. Discussion

The ability to pump energy into the magnetic signal during its propagation is the most appealing property of the described interconnects. The pumping is via the magneto-electric coupling in the multiferroic, where the some portion of the electric energy provided to the capacitor is transferred to the energy of the magnetic signal. The amplitude of the magnetic signal (i.e. the angle of magnetization rotation) is controlled by the applied voltage and saturates to the certain value as the electric field across the piezoelectric reaches its steady state distribution. This property is critically important for logic circuit construction allowing us to minimize the effect of structure imperfections and to make logic circuit immune to the thermal noise. It should be also noted that the absolute value of magnetization change in the interconnect may exceed the initial magnetization state of the sender nano-magnet. For instance, the Y component of magnetization of the nano-magnet A may be 0.1Ms, while the Y magnetization of the magnetic signal in the interconnect may saturate around 0.5 Ms as illustrated by numerical modeling in the previous Section. In other words, the proposed interconnects may serve as an amplifier for magnetic signals similar to the multiferroic spin wave amplifier described in Ref.[15]. Another important property of the proposed interconnect is the ability to control the direction of signal propagation by the applied voltage. Similar to the "All Spin Logic" approach [3], where the direction magnetic signal is defined by the direction of spin polarized current flow, the change of magnetization in the multiferroic interconnect follows the charge diffusion. This property resolves the problem of input-output isolation and provides an additional degree of freedom for logic circuit construction.

Energy dissipation in a two-phase magnetoelastic/piezoelectric multiferroic has been studied in Refs.[13,16,17]. According to the estimates, a single two-phase magnetoelastic/piezoelectric multiferroic single-domain shape-anisotropic nano-magnet can be switched consuming as low as 45kT for a delay of 100ns at room temperature, where the main contribution to the dissipated energy comes from the losses during the charging /discharging ($\approx CV^2$) [17]. The capacitance of a micrometer long multiferroic interconnects comprising 40nm of PZT and 4nm of Ni with the width of 40nm is about 15fF, and the control voltage required for 90 degree anisotropy easy-axis change is 0.6MV/m×40nm=24mV. Thus, assuming all the electric energy dissipated during signal propagation one has 9aJ per signal per 1μm transmitted. It is important to note, that according to the theoretical estimates [17], the energy dissipation increases sub-linearly with the switching speed. For example, in order to increase the switching speed by a factor of 10, the dissipation needs to increase by a factor of 1.6.

The propagation of magnetization signal involves several physical processes: charge diffusion, mechanical response of the piezoelectric to the applied electric field, change of the anisotropy field caused by the stress, and magnetization relaxation. Thus, the total delay time $\tau_t$ is the sum of the following:

$$\tau_t = \tau_e + \tau_{mech} + \tau_{mag}, \qquad (5)$$

where $\tau_e$ is the time delay due to the charge diffusion $\tau_e$=RC; $\tau_{mech}$ is the delay time of the mechanical response $\tau_{mech} \approx d/v_a$, where d is the thickness of the piezoelectric layer, $v_a$ is the speed of sound in the piezoelectric, $\tau_{mag}$ is the time required for the spins of magnetostrictive material to follow the changing anisotropy field. In the theoretical model presented in the previous Section, we introduced a direct coupling among the electric field and the anisotropy field (i.e. Eq.4) presuming an immediate anisotropy field response on the applied electric field. The latter may be valid for the thin piezoelectric layers (e.g. taking $d$=40nm, $v_a$=1×10$^3$m/s, $\tau_{mech}$ is about 40ps). We also introduced a high damping coefficient $\eta$, which minimizes the magnetic relaxation time $\tau_{maa}$<50ps. In this approximation, the speed of signal propagation is mainly defined by the charge diffusion rate. The smaller is RC is the faster is the charge diffusion and the lower are the energy losses for interconnect charging/discharging.

In Figure 3, we show the results of numerical modeling on the speed of signal propagation for different thicknesses of piezoelectric layer. The four curves correspond to the signal propagation in the interconnects with different PMN-PT thickness (e.g. 20nm, 40nm, 80nm, and 200nm) respectively. The thickness of nickel layer is 4nm for all cases. We also plotted a reference line corresponding to the magnetostatic spin wave with typical group velocity of 3.1×10$^4$ m/s. According to these estimates, one may observe that the magnetic signal in the multiferroic interconnect may propagate faster than the spin wave at short distances (<500nm) and slower than the spin wave at longer distances. The latter leads to an interesting question whether or not it is possible to transmit magnetic signals faster than the spin wave in the magnetoelastic material. Though magnetic coupling does not define the speed of signal propagation, it should determine the trajectory of spin relaxation. Exceeding the speed of spin wave in ferromagnetic material may lead to a chaotic magnetic reorientation along the ferromagnetic layer. This is a one of many questions to be answered with further study.

Finally, we want to compare the main characteristics of different magnetic interconnects and discuss their advantages and shortcomings. Moving a domain wall is a reliable and experimentally proven way for magnetic signal transmission [18]. A domain wall propagates through a magnetic wire as long as an electric current or an external magnetic field is applied, and remains at a constant position if the driving force is absent. This property is extremely useful for building magnetic memory (e.g. the "racetrack" memory [19]). The speed of domain motion may exceed hundreds of meters per second if the driving electric current has a sufficiently large density (e.g. 250m/s at 1.5×10$^8$ A/cm$^2$ from Ref.[19]). Slow propagation speed and high energy per bit are the main disadvantages of the logic circuits exploding domain wall motion.

Relatively faster and less power consuming are the interconnects made as the sequence of nano-magnets, where the nearest neighbor nano-magnets are coupled via the dipole-dipole interaction (so called the Nano-Magnetic Logic (NML) [20]). Experimentally realized wires formed from a line of anti-ferromagnetically ordered nano-magnets show signal propagation speed up to 10$^3$m/s with an internal (without the losses in the magnetic field generating contours) power dissipation per bit about tens of atto Joules[14]. There is a tradeoff between the speed of signal propagation and the dissipated energy. The slow is the speed of propagation and lower is the energy dissipated within the interconnect. The main shortcoming of the nano-magnet interconnect is associated with reliability, as the thermal noise and fabrication-related imperfections can cause errors in signal transmission and overall logic functionality of the NML circuits [21].

Interconnects exploiting spin waves may provide signal propagation with the speed of $10^4$m/s-$10^5$m/s . At the same time, the amplitude of the spin wave signal is limited by the several degrees of magnetization rotation in contrast to the complete magnetization reversal provided by the domain wall motion or NML. The amplitude of the spin wave decreases during the propagation (e.g. the attenuation time for magnetostatic surface spin waves in NiFe is 0.8ns at Room temperature [22]). The unique advantage of the spin wave approach is that the interconnects themselves can be used passive logic elements exploiting spin wave interference. The latter offers an additional degree of freedom for logic gate construction and makes it possible to minimize the number of nano-magnets per logic circuit [4].

The All Spin Logic (ASL) proposal suggests the use of spin polarized currents for nano-magnet coupling [3]. This approach allows for much higher defect tolerance as the variations in the size and the position of input/output nano-magnets are of minor importance. It is also scalable since shorter distances between the input/output cells would require less spin polarized currents for switching. According to the theoretical estimates [23], ASL can potentially reduce the switching energy-delay product. The major constrain is associated with the need of the spin-coherent channel, where the length of the interconnects exploiting spin-polarized currents is limited by the spin diffusion length.

The described magnetic interconnects based on synthetic multiferroics combines high transmission speed (as fast as the spin waves) with the possibility of transmitting large amplitude signals (up to 90 degrees of the magnetization rotation). As we stated above, the main appealing property of the proposed interconnect is the ability to keep the constant amplitude of the magnetization signal. All these advantages are the result of using the electro-magnetic coupling in multiferroics allowing us to pump energy from the electric to the magnetic domain. Based on the presented estimates, the energy per transmitted bit may be as low as several atto Joules per 100nm transmitted distance. From the practical point of view, the implementation of synthetic multiferroic interconnects is feasible, as it relies on the integration of the well-known materials (e.g. PMN-PT and Ni) and can be integrated on a silicon platform. However, the dynamics of the electro-mechanical-magnetic coupling in synthetic multiferroics remains mainly unexplored. The expected challenges are associated with the limited scalability, as the thickness of the piezoelectric should be sufficient to generate stress required for anisotropy change. In Table I, we have summarized the estimates on the main characteristics of different magnetic interconnects and outlined their major advantages and shortcomings.

### V. Conclusions

In summary, we considered a novel type of magnetic interconnect exploiting electro-magnetic coupling in two-phase synthetic multiferroics. According to the presented estimates, synthetic multiferroic interconnects combines the advantages of fast signal propagation (up to $10^5$m/s) and low power dissipation (less than 1aJ per 100nm). The most appealing property of the multiferroic interconnects is the ability to pump energy into the magnetic signal and amplify it during propagation. Voltage-driven magnetic interconnect may be utilized in nano-magnetic logic circuitry and provide an efficient tool for logic gate construction. The fundamental limits and practical constrains inherent to two-phase multiferroics are associated with the efficiency of the stress-mediated coupling at high frequencies. There are many questions related to the dynamic of the stress-mediated signal propagation, which will be clarified with further theoretical and experimental study.

**Acknowledgments**

This work was supported in part by the FAME Center, one of six centers of STARnet, a Semiconductor Research Corporation program sponsored by MARCO and DARPA and by the National Science Foundation under the NEB2020 Grant ECCS-1124714.

**Figure Captions**

Figure 1(A) Schematics of the synthetic multiferroic interconnect comprising a piezoelectric layer (PMN-PT) and a magnetostrictive layer (Ni). The structure resembles a parallel plate capacitor. An applying of voltage at point A results in the charge diffusion through the plates. In turn, an electric field applied across the piezoelectric produces stress, which rotates the easy axis of the magnetoelastic material. (B) The equivalent electric circuit – RC line, which is used in numerical simulations. (C) Results of numerical simulations showing the distribution of the electric field and the magnetization along the interconnect. The change of magnetization in the magnetoelastic layer follows the charge diffusion.

Figure 2. Results of numerical modeling showing the normalized magnetization $M_Y/M_S$ as a function of time. The two sets of curves show magnetization trajectories following the initial state of the sender nano-magnet A (e.g. along or opposite to axis Y). The black, the red, and the blue curves show magnetization at 1.0µm, 2.0µm, and 3.0µm distance away from the starting point A.

Figure 3. Results of numerical modeling illustrating the speed of signal propagation in the synthetic multiferroic interconnect. There are shown several curves corresponding to different thickness of the PMN-PT layer (20nm, 40nm, 80nm, and 200nm). The blue line is the reference data for the Magnetostatic Surface Spin Wave (MSSW) with group velocity of $3.0\times10^4$ m/s.

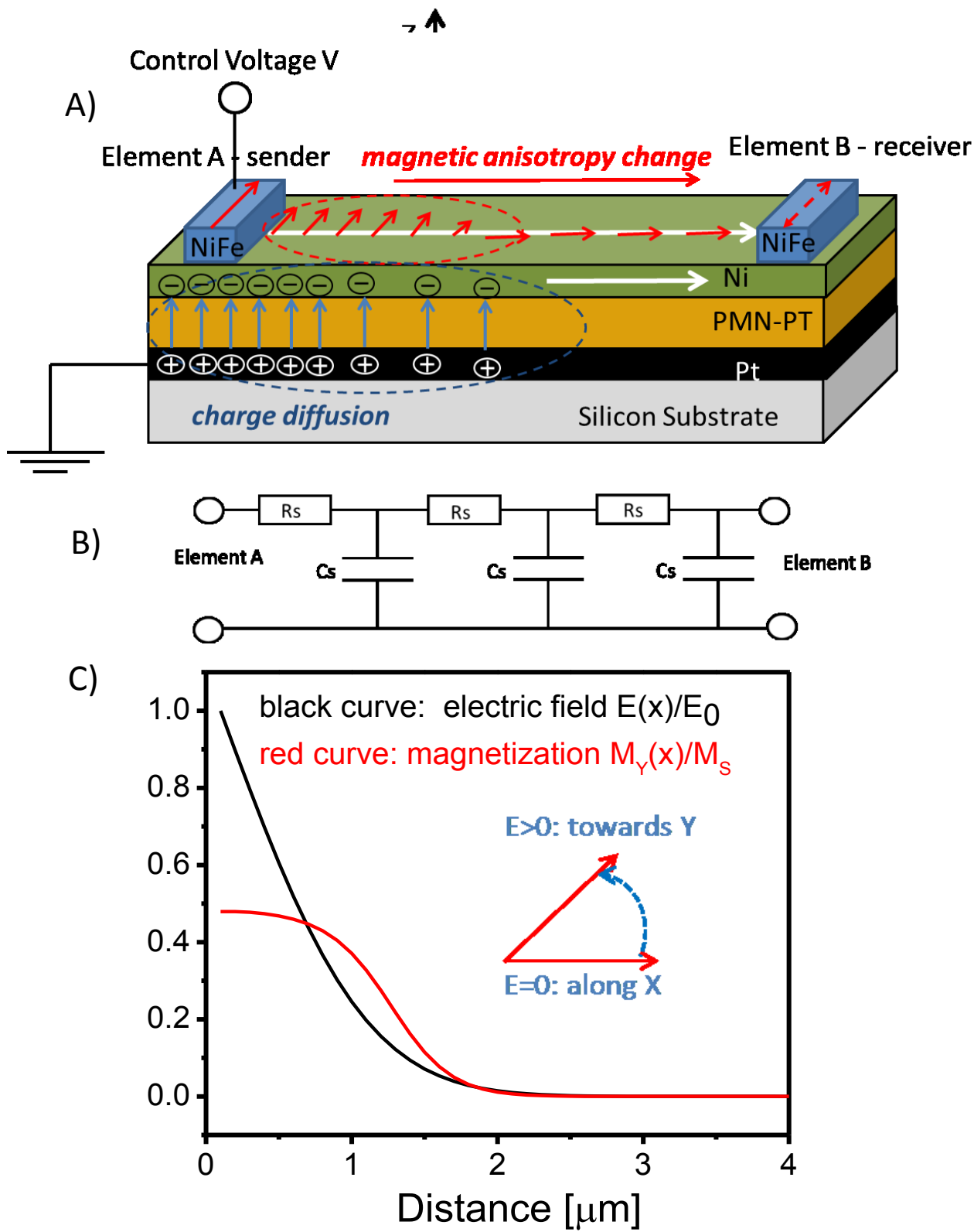

**Figure 1**

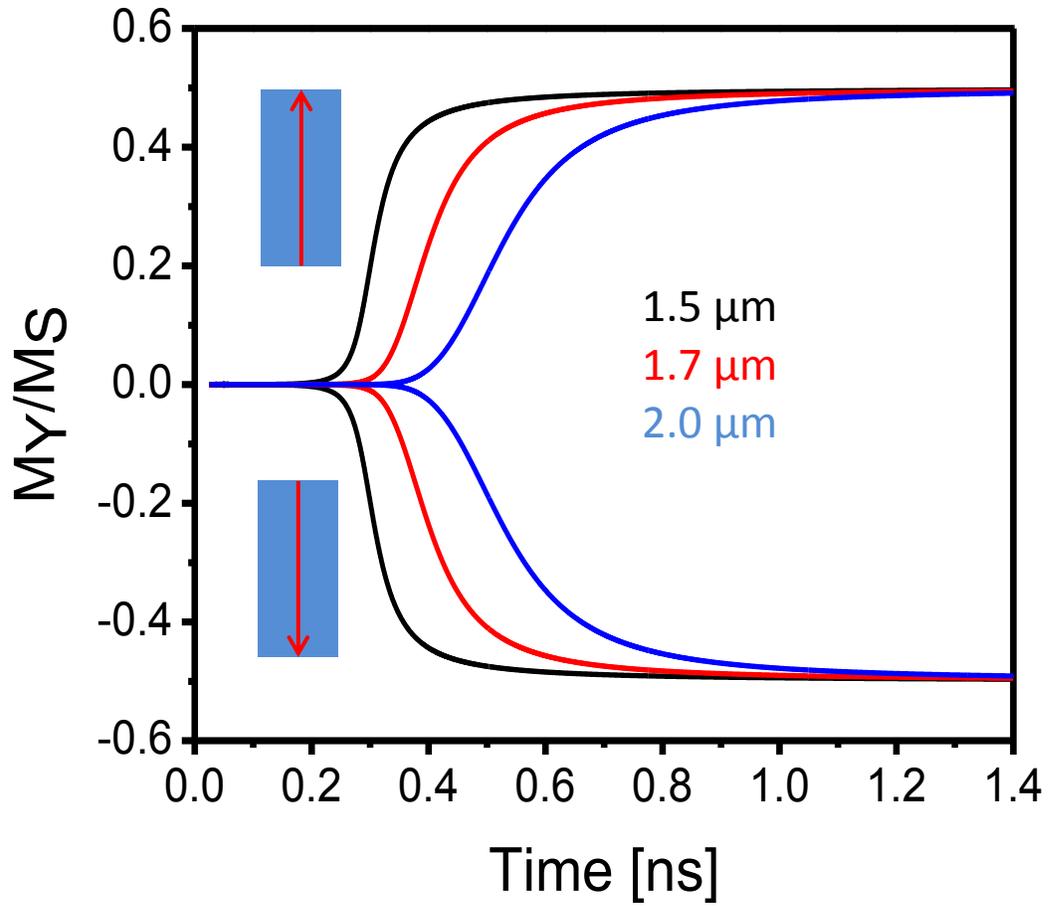

**Figure 2**

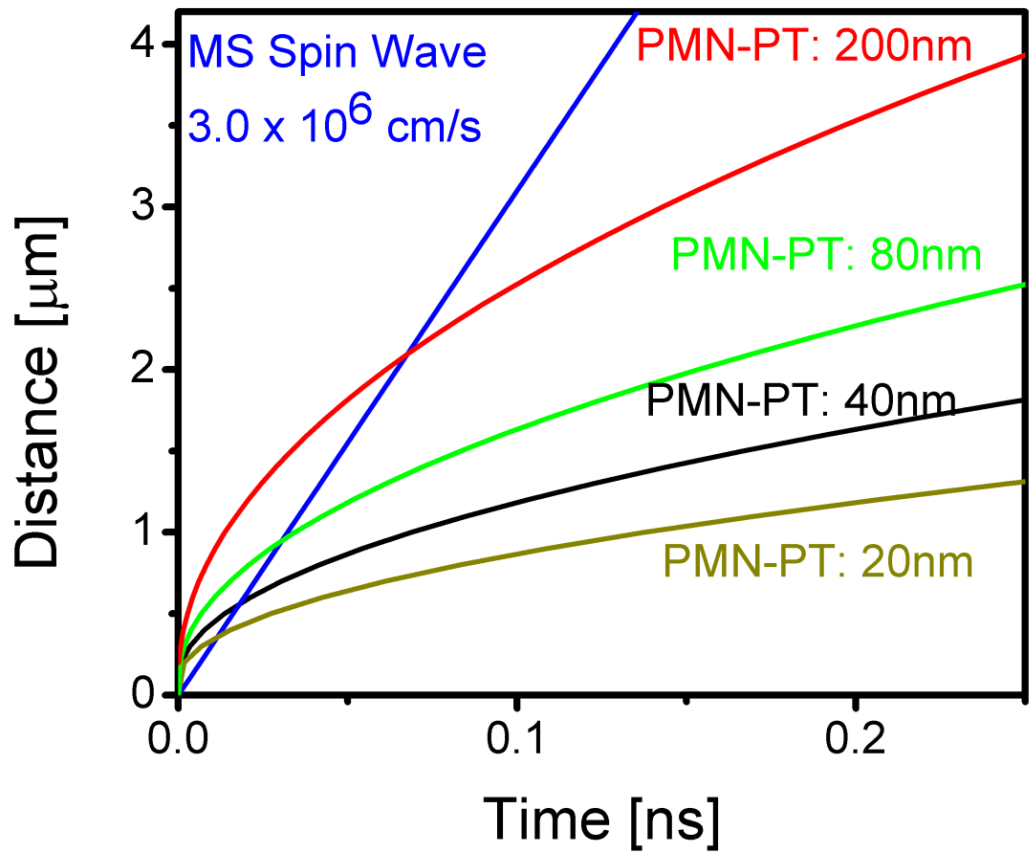

**Figure 3**

**Table I**

|  | **Domain Wall** | **MCA** | **Spin Wave** | **ASL** | **Multiferroics** |
|---|---|---|---|---|---|
| **Mechanism of coupling** | Domain wall motion | Dipole-dipole coupling | Spin waves | Spin polarized current | Magnetization signal |
| **Speed of propagation** | $10^2$ m/s | $10^3$ m/s | $10^4$m/s-$10^5$m/s | * $10^5$m/s | * $10^5$m/s |
| **Energy dissipated per bit transmitted** | >1000 aJ | **1aJ | 0.1aJ | N/A | 1aJ |
| **Main advantage** | Non-volatile, can be stopped at any time and preserve its position | Internal dissipated energy approaches zero at the adiabatic switching | Computation in wires – additional functionality via wave interference | Scalable, defect tolerant | Fast signal propagation, signal amplification |
| **Main disadvantage** | Slow and energy consuming | Effect of thermal noise increases with the propagation distance | Propagation distance is limited due to the spin wave damping | Propagation distance is limited by the spin diffusion length | Limited scalability |

* Signal propagation speed is determined by the charge diffusion and decreases with the distance.
** The estimates for $10^3$m/s propagation speed and include only for the energy dissipated inside the magnetic interconnect (without considering the energy losses in the magnetic field generating contours)